\documentclass[traditabstract]{aa} % for the abstract without structuration 
\usepackage{graphicx}
\usepackage{txfonts}
\def\gsim{ \lower .75ex \hbox{$\sim$} \llap{\raise .27ex \hbox{$>$}} } 
\def\lsim{ \lower .75ex\hbox{$\sim$} \llap{\raise .27ex \hbox{$<$}} }

\def\ama{$E_{\rm peak}-E_{\rm iso}$}
\def\yone{$E_{\rm peak}-L_{\rm iso}$}
\def\yonet{$E_{\rm peak}^{t}-L_{\rm iso}^{t}$}

\def\sw{{\it Swift}}
\def\sax{{\it Beppo}SAX}
\def\he{{\it Hete-II}}
\def\ko{{\it Konus-Wind}}
\def\ba{BATSE}
\def\fermi{{\it Fermi}}

\def\ep{$E_{\rm peak}$}
\def\epo{$E^{\rm obs}_{\rm peak}$}
\def\epof{$E^{\rm obs}_{\rm peak}$--$F$}
\def\epop{$E^{\rm obs}_{\rm peak}$--$P$}
\def\eiso{$E_{\rm iso}$}
\def\liso{$L_{\rm iso}$}
\def\eg{$E_{\rm \gamma}$}

\begin{document}

\title{Spectral--Luminosity relation within individual {\it Fermi} GRBs}
%\subtitle{Time dependent \yone\ correlation}

\author{
G. Ghirlanda\inst{1},  L. Nava\inst{1,2} \and G. Ghisellini\inst{1}
}
\institute{INAF - Osservatorio Astronomico di Brera, Via E. Bianchi 46, I--23807 Merate, Italy \\ 
\email{giancarlo.ghirlanda@brera.inaf.it} 
\and Universit\`a degli Studi dell'Insubria, via Valleggio 11, I--22100 Como, Italy}

%  \date{Received ....; accepted ....}

% \abstract{}{}{}{}{} 
% 5 {} token are mandatory
 
\abstract{
We study the spectra of all long Gamma Ray Bursts (GRBs) of known redshift detected by the {\it Fermi} satellite.
Their fluxes and fluences are large enough to allow a time dependent study of their spectral characteristics in the 8 keV-1 MeV energy range. We find that the peak energy $E_{\rm peak}$ of their $EL(E)$ spectrum correlates with the luminosity in a remarkable tight way within individual bursts.
This time resolved \yone\ correlation is very similar for all the considered bursts, and has a slope  and normalisation similar to the analogous \yone\ correlation defined by the time integrated spectra of different bursts detected by several different satellites. For a few of the considered GRBs, we could also study the behaviour of the \yone\ correlation during the rising and decaying phases of individual pulses within each burst, finding no differences.
Our results indicate the presence of a similar physical mechanism, operating for the duration of different GRBs, 
linking tightly the burst luminosity with the peak energy of the spectrum emitted at different times. Such a physical 
mechanism is the same during the rise and decay phase of individual pulses composing a GRB. These results,
while calling for a robust physical interpretation, strongly indicate that the \yone\ spectral energy correlation found 
considering the time integrated spectra of different bursts is real, and not the result of instrumental selection effects.
}
\maketitle

%________________________________________________________________

\section{Introduction}

One of the key properties of the prompt emission of Gamma Ray Bursts
(GRBs) that is still poorly understood concerns the spectral--energy
correlations found considering the time integrated spectra of bursts
of known redshift, and for which we can define the peak energy of
the spectrum, \ep, in a $EL(E)$ representation.
\ep\ correlates with the isotropic luminosity \liso\ (Yonetoku et al. 2004), with the isotropic 
energy \eiso\ (Amati et al. 2002) and more tightly with the collimation--corrected energy \eg\
(Ghirlanda, Ghisellini \& Lazzati 2004). 
There are two very strong motivations to study these correlations: one is to understand their  
physical origin, that can disclose a still not understood basic property of GRBs 
(Yamazaki, Ioka \& Nakamura 2004; 
Lamb, Donaghy \& Graziani 2005;
Rees \& Meszaros 2005; 
Levinson \& Eichler 2005; 
Toma et al. 2005; 
Eichler \& Levinson 2006, 2004; 
Barbiellini et al. 2006;
Thompson 2006; 
Ryde et al. 2006; 
Giannios \& Spruit 2007; 
Thompson, Meszaros \& Rees, 2007; 
Guida et al. 2008; 
Panaitescu 2009)
and the other is the possibility to use these correlations to
standardise the GRB energetics, making them cosmological tools
(Ghirlanda et al. 2004a, 2006, 2006a; 
Firmani et al. 2005, 2006, 2007; 
Xu, Dai \& Liang 2005; 
Liang \& Zhang 2005, 2006; 
Wang \& Dai 2006;
Qi, Wang \& Lu 2008;
Li et al. 2008;
Liang et al. 2008).

The debate about the reality of these correlations is hot,
since some authors pointed out that they can be the result of 
observational selection effects (Nakar \& Piran 2005; Band \& Preece 2005; Butler et al. 2007, Butler, Kocevski \& Bloom 2009; Shahmoradi \& Nemiroff 2009)
while others argue that selection effects, 
even if surely present, play a marginal role 
(Ghirlanda et al. 2005, Bosnjak et al. 2008, Ghirlanda et al. 2008; Nava et al., 2008; Krimm et al. 2009; Amati et al. 2009).

One possibility to get some insight on this issue is to study
{\it individual}, bright bursts to see if, during the prompt phase,
the luminosity and peak energy at different times correlate.
If they do, and furthermore if the slope of this {\it time resolved}  correlation (indicated \yonet\ hereafter)
is similar to the {\it time integrated} \yone\ correlation found 
%using the time integrated spectra of different
among different bursts, then we should conclude that the spectral energy correlations
are surely a manifestation of the physics of GRBs, and are not the result
of instrumental selection effects.

%_____________________________________________________________
%                                             Two column Table 
%_____________________________________________________________
%
\begin{table*}
\caption{\fermi\ long GRBs with redshift. We mark in boldface the GRBs with a detection also in the LAT instrument on--board \fermi.* The energy peak flux of GRB 080916C in the 20keV--10MeV range is from Golenetskii et al. 2008}
\label{tab1}      
\centering          
\begin{tabular}{l l l l l l l l l l l}     % 7 columns 
\hline\hline       
                      % To combine 4 columns into a single one G
   GRB    &$z$    & $\alpha$ & $E_{\rm peak}$ & $\beta$ & $P$  & $F_{-6}$ &range &GCN   &$L_{\rm iso,52}$  & $E_{\rm iso,52}$ \\
          &       &          &      keV       &        & ph/s/cm$^2$ &erg/cm$^{2}$ &keV & number  &erg/s    &erg     \\ 
\hline                    
080810(549)       &3.35  &--0.91$\pm$0.12 &313.5$\pm$3.6 &                &1.85$\pm$0.16 &6.9$\pm$0.5    &50--300  &8100 &7.84  &33.2 \\
080905(705)       &2.374 &--1.75$\pm$0.12 &              &                &0.21$\pm$0.02 &0.04$\pm$0.003 &20--1000 &8205 &      &  \\
080916(406)       &0.689 &--0.9$\pm$0.1   &109$\pm$9     &                &4.5$\pm$0.7   &15$\pm$5       &25--1000 &8263 &0.142 &2.25 \\
{\bf 080916(009)} &4.35  &--0.91$\pm$0.02 &424$\pm$24    &--2.08$\pm$0.06 &   1.2e-5$*$           &190            &8--30000 &8278 &190   &563  \\
080928(628)       &1.692 &--1.80$\pm$0.08 &              &                &              &1.5$\pm$0.1    &50--300  &8316 &      &     \\
081007            &0.529 &--1.4$\pm$0.4   &40$\pm$10     &                &2.2$\pm$0.2   &1.2$\pm$0.1    &25--900  &8369 &0.041 &0.172 \\
081222(204)       &2.77  &--0.55$\pm$0.07 &134$\pm$9     &--2.10$\pm$0.06 &14.8$\pm$1.4  &13.5$\pm$0.8   &8--1000  &8715 &20.6  &35.4 \\
{\bf 090323(002)} &3.57  &--0.89$\pm$0.03 &697$\pm$51    &                &12.3$\pm$0.4  &100$\pm$1      &8--1000  &9035 &47.2  &338 \\
{\bf 090328(401)} &0.736 &--0.93$\pm$0.02 &653$\pm$45    &--2.2$\pm$0.1   &18.5$\pm$0.5  &80.9$\pm$1     &8--1000  &9057 &1.96  &21.2 \\
090423(330)       &8.2   &--0.77$\pm$0.35 &82$\pm$15     & 		  &3.3$\pm$0.5   &1.1$\pm$0.3    &8--1000  &9229 &18.8  &10.2  \\
090424(592)       &0.544 &--0.90$\pm$0.02 &177$\pm$3     &--2.9$\pm$0.1   &137$\pm$5     &52$\pm$1       &8--1000  &9230 &2.12  &4.48  \\
090618(353)       &0.54  &--1.26$\pm$0.04 &155.5$\pm$11  &--2.5$\pm$0.25  &73.4$\pm$2.0  &270$\pm$6      &8--1000  &9535 & 1.0  &25.7  \\
\hline                  
\end{tabular}
\end{table*}
%----------------------------------------------------------------------------------------------------

Some attempts have already been done. Liang, Dai \& Wu (2004) considered \ba\ bursts without 
known redshifts and showed the presence of a correlation between the (observer frame) peak energy and the 
flux within individual bursts which they interpret as suggestive of a physical origin of the \yone\ correlation holding among 
the GRBs with measured redshift. However, to compare direcly the \yonet\ correlations of individual GRBs  with the \yone\ correlation defined with time integrated spectra it is necessary to know the redshift (which is instead unknown for most of \ba\ bursts).  Recently, Firmani et al. (2009) considered \sw\ bursts of known redshift finding a rather strong \yonet\ correlation within individual GRBs. Having the redshift, they could directly compare the time resolved correlation of  different bursts, finding that the ensemble of data points in the \yone\ plane shows a  correlation similar to that defined with the time integrated spectra of the same burst sample. The Burst Alert Telescope (BAT) onboard \sw, however, with its limited energy range (15--150 keV), is not particularly suited for GRB spectral analysis,  especially when dealing with time resolved spectra. To overcome this limitation, Ohno et al. (2009) combined the \sw--BAT and {\it Suzaku}--WAM spectral data to study the spectral evolution of GRB 061007 and investigate the time evolution of the \yone\ correlation within the two pulses of this burst. They found that the time resolved pulses also satisfy the \yone\ correlation defined by time integrated spectra. 
%although this is more evident in the second pulse of GRB 061007.  Note that the first pulse of GRB 061007 could be regarded as a ``precursor'' of the main event (Burlon et al. 2008) and %that the spectral evolution of precursors and of the corresponding main pulses can follow parallel tracks in the observer frame plane \epop\ (where P is the flux) although with misaligned %normalizations (Burlon et al. 2009). 
A more systematic analysis of the time resolved spectral properties of \sw--{\it Suzaku} GRBs (Krimm et al. 2009) shows that individual pulses within a GRB are consistent with the \ama\ correlation defined by the time averaged spectra. They consider the spectra integrated over the duration of individual pulses. In this case \eiso\ is computed on different integration timescales. They find that the pulses follow a correlation parallel to the \ama\ defined with time integrated spectra, but with a higher normalization. Instead, the comparison of time resolved spectra with the \yone\ correlation is independent of the spectral integration time since this correlation involves the luminosity \liso\ rather than the energy \eiso.  It is therefore important to study the presence of a \yone\ correlation (as done by Ohno et al. 2009 for a single event) by concentrating on GRBs with measured redshift, in oder to compare their time resolved \yonet\ correlation with that defined with the time integrated spectra. In particular we aim at studying how single GRBs evolve in the \yone\ plane rather than considering them globally (as in Firmani et al. 2009). Furthermore, we would like to study the rise and decay phases of individual pulses.

The Gamma ray Burst Monitor (GBM, Meegan et al. 2009) onboard the \fermi\ satellite  covers a wide energy range (8 keV -- 30 MeV) and, although slightly less sensitive than \sw/BAT, it is better for studying the spectral 
properties of the prompt emission of GRBs. In addition, the Large Area Telescope (LAT) sensitive in the 0.1--100 GeV
energy range can complement the spectral information for the few bursts it can detect.
In one year of operation (up to the end of July 2009),  \fermi/GBM detected about 200 bursts
and for about  half of them the (time integrated) spectral analysis returned a well defined \epo.

In this paper we study \fermi\ GRBs, selecting those of known redshift, to be able to compare their different 
evolutionary tracks in the \yone\ plane with the correlation defined by the time integrated spectra (see e.g. Ghirlanda et al. 2009 for a recent compilation of the \yone\ correlation). For the brightest bursts we study if the rising and decaying phases
of individual pulses behave differently in the \yone\ plane, since this can give
important clues for our physical understanding of the emission mechanism operating during the GRB prompt phase.

The paper is organised as follows:
in \S 2 we describe our \fermi\ GRB sample, whose time--integrated 
spectral properties are presented in \S 3 and compared to the \ama\ and \yone\ correlation 
defined by pre--\fermi\ GRBs.
In \S 4 we describe the time resolved spectral analysis
whose results are given in \S 5. In \S 6 we discuss our findings and we draw our conclusions.
A standard cosmology for a flat universe with $h_{0}=\Omega_{\Lambda}=0.7$ is assumed.

\section{The sample}

We considered the GRBs detected by the GBM (up to the end of July 2009) of known redshift.
They are 13 events. Among these GRB 090510 ($z$=0.903, Rau et al. 2009) is a short burst, 
having an observer frame duration of less than 2 s, and will not be considered here.

Tab. \ref{tab1} lists the 12 long GRBs, their time integrated spectral parameters (Col. 3 to Col. 8)
and the derived isotropic luminosity (\liso, Col.~10) and isotropic energy 
(\eiso, Col.~11), computed in the rest frame 1 keV -- 10 MeV energy range. 
The spectral parameters in Tab. \ref{tab1} have been collected from the literature 
(references are given Col. 9): they were obtained through the analysis of the time integrated 
spectrum extracted from the GBM data.
In two cases (GRB 080905 -- Bhat et al. 2008; and GRB 080928 -- Paciesas et al., 2008) 
the time integrated spectrum is fitted by a single power law and, therefore, the peak energy is unconstrained.
In five cases the time integrated spectrum is modeled with a power law 
ending with an exponential cutoff at high energies. In the remaining five cases it is modeled by a Band function. 

Two of the three GRBs detected by the LAT (in boldface in Tab. \ref{tab1})  belong to the latter group 
(i.e GRB 080916C, Tajima et al., 2008; Abdo et al. 2009; and 
GRB 090328, Cutini et al., 2009). These bursts show a high energy power law component and their 
observed peak energies \epo\ are the largest of the sample. Note that Tab. \ref{tab1} lists also the most distant burst: 
GRB 090423, with $z$=8.2 (Tanvir et al., 2009).

\section{Time integrated spectra: the \ama\ and the \yone\ correlations}

\begin{figure}
\includegraphics[width=9cm]{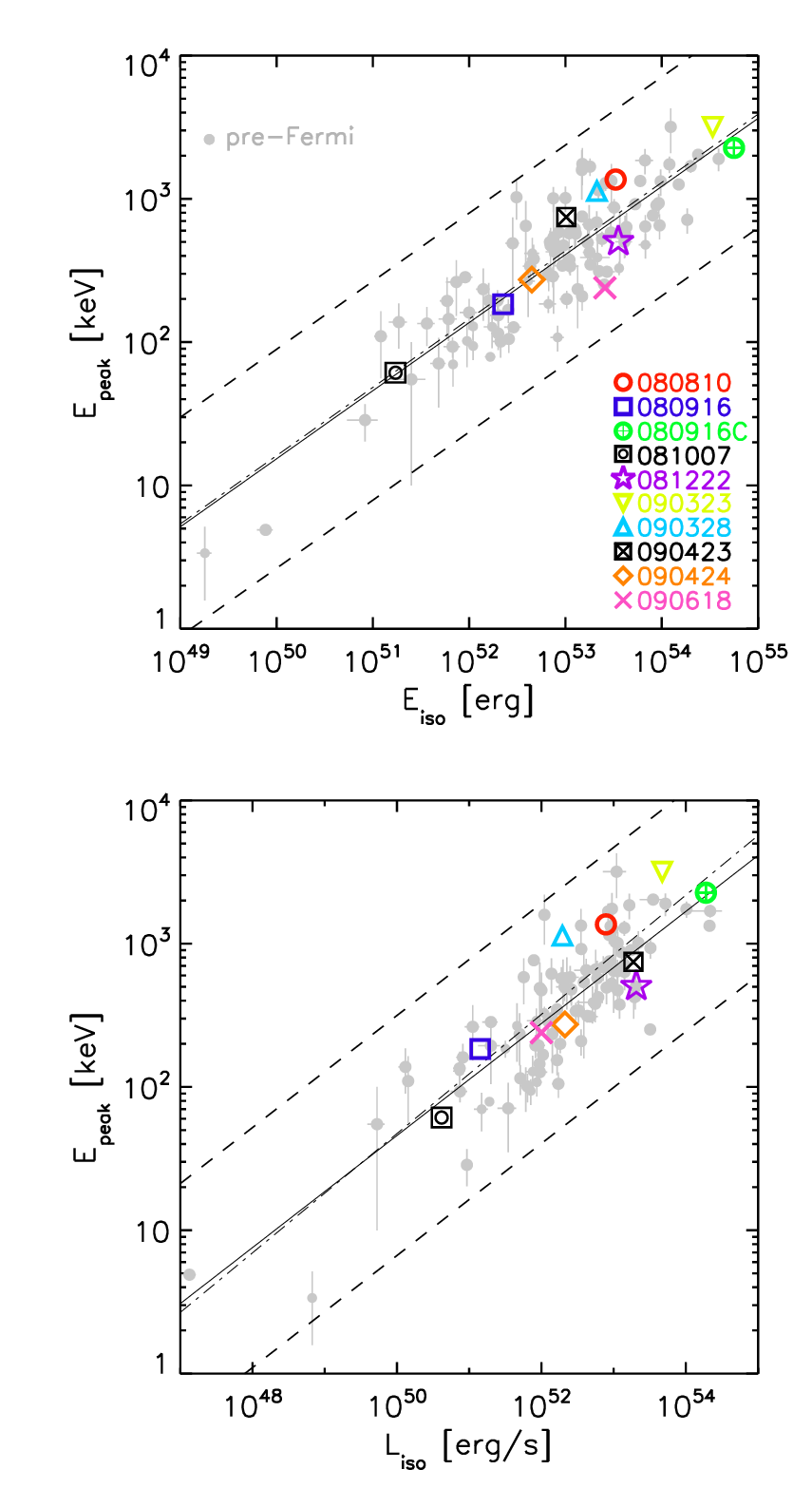}
\caption{
The 10 \fermi--GRBs with redshift and \epo\ in the \ama\ (top panel) and \yone\ 
(bottom panel) planes. 
Their position is compared with the 
correlations defined by the sample of 100 GRBs detected by other 
instruments (grey filled circles), i.e. pre-\fermi\ sample. 
The solid line is the best fit to the pre-\fermi\ sample, 
while the dashed lines represent its 3$\sigma$ scatter. The dot--dashed line 
is the best fit to the 10 \fermi\--GRBs. 
\fermi\ bursts appear to be fully consistent both with the \ama\ 
and the \yone\ correlations.
}
\label{corr_upd}
\end{figure}

First, we check the consistency of \fermi\ bursts with the \ama\ and the \yone\ correlations defined 
by the time integrated spectra of GRBs detected by other instruments. The most updated pre--\fermi\ sample of GRBs with  known $z$ and \epo\ contains 100 objects
detected by different instruments. Fig. \ref{corr_upd} shows these bursts (grey filled circles) in the 
\ama\ and \yone\ planes. In both planes they define a strong correlation (the probability that the correlation is by chance is reported in Col. 4 of Table \ref{tab2}), confirming recent analysis 
(Nava et al. 2008; Ghirlanda et al. 2009). The slope and normalization of the fit of these correlations with a power law  are reported in Table \ref{tab2}. 
To show where the \fermi\ bursts lie in these planes we estimated \eiso\ and \liso\ using the spectral parameters reported 
in the literature and listed in Table \ref{tab1}, excluding the two GRBs fitted with a single power law (i.e. with an unconstrained \epo).
Fig. \ref{corr_upd} shows that the position of \fermi\ bursts is consistent with both correlations. We have fitted the 100 pre--\fermi\ GRBs, the 10 \fermi\  GRBs and the combined sample of  110 GRBs with the least square method. The best fit spectral parameters (normalization $K$ and slope $\delta$) and the probability that the correlation is by chance ($P$) are reported in Table \ref{tab2}.
% ------------------------------------------------------------------------ 
%----------------------------------------------------------------------------

% ---------------------------------------------------------------------------------------------------------- 
\begin{figure*}
\centering
\includegraphics{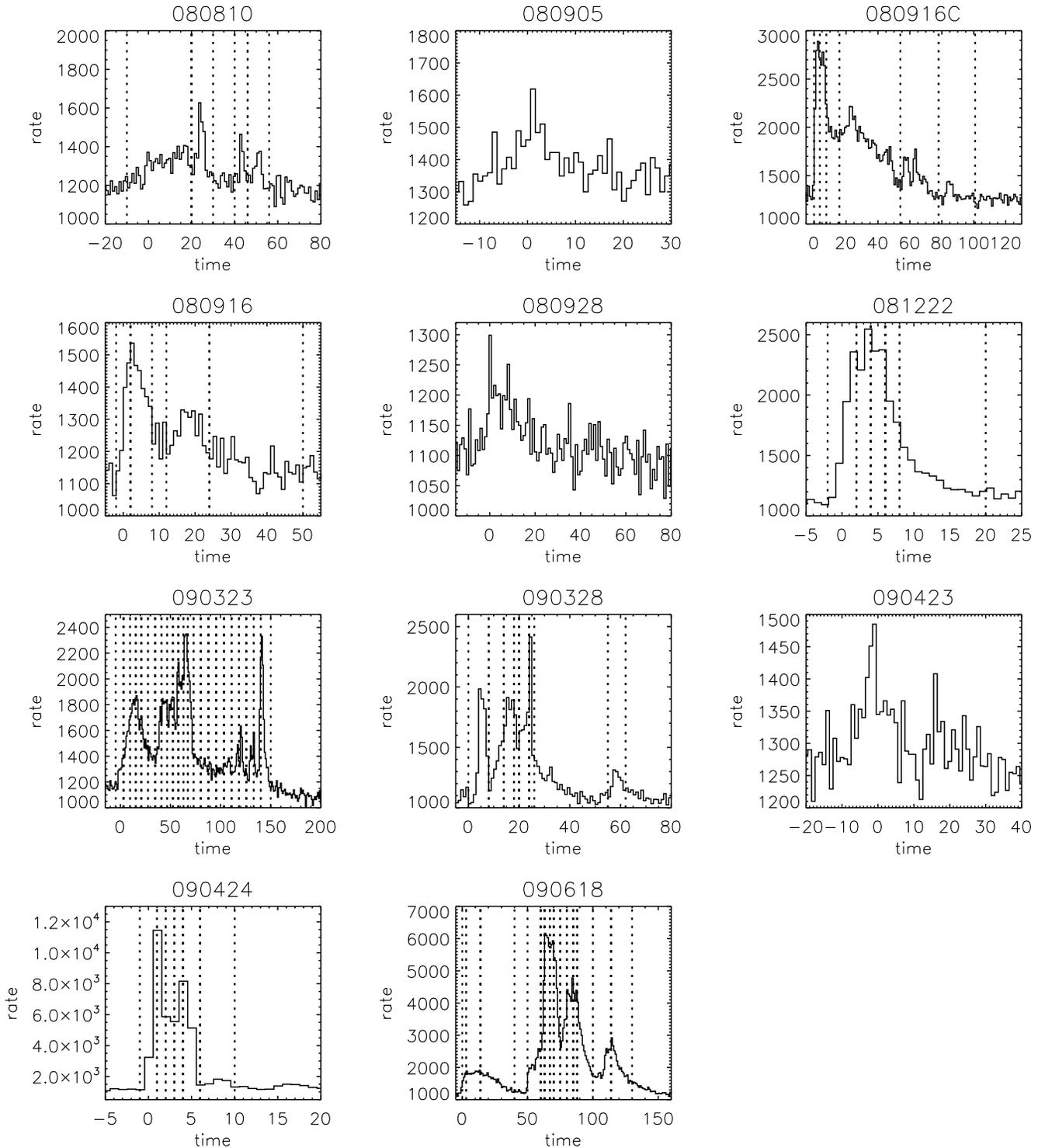}
\caption{
\fermi\ light curves of the bursts with 
known redshift detected by the GBM from August 2008 to July 2009. 
Light curves have time resolution of 1 s and are not 
background subtracted. The vertical dotted lines mark the 
time intervals selected for the extraction of the time 
resolved spectra. The light curve of GRB 081007 is not 
reported because the data of this burst could not be found. 
}
\label{lc}
\end{figure*}
%-----------------------------------------------------------------------------------------------------------

\section{Time resolved spectra: data analysis}

The data of GRBs detected by the GBM since August 2008 are publicly available
at {\tt http://fermi.gsfc.nasa.gov/}. 
The GBM consists of 12 NaI and 2 BGO scintillation detectors 
differently oriented so to derive the GRB position through the 
comparison of the count rates of the different detectors. 
The NaI cover the low energy spectral domain from 8 keV to $\sim$1 
MeV while the BGO detector (a factor 10 thicker than the NaI) are 
sensitive in the 0.2--30 MeV energy range. 
The GBM uses several trigger algorithms to detect a GRB. 
These are determined by 
different choices of the integration timescales and energy range. 
A trigger algorithm similar to that of BATSE Large Area Detectors (LAD)
 operates in the 50--300 keV band with a minimum significance of 5.4$\sigma$ 
for the detection of the GRB over the background 
(which has typically a rate of 300--350 counts/s in this range -- e.g. 
Meegan et al. 2009) on the 16 ms--4 s timescales. 
An event is flagged as a burst if at least two NaI detectors 
are simultaneously triggered.

% For each GRB a directory containing the relevant data is available. 
For our analysis we considered the TTE (Time Tagged Event) files containing 
the counts in 128 energy channels relative to the burst period. 
We considered the TTE files of the two NaI detectors 
triggered by the GRB. From these we extracted the GRB light 
curve and the time resolved spectra with the {\it gtbin} tool 
(as part of the \texttt{ScienceTools-v9r8p2-fssc-20090225}). 
Light curves were extracted by summing the count rates over 
the 8 keV--1 MeV energy range of the NaI detectors that were triggered and the 
200 keV--30 MeV energy range of the 2 BGO detectors. 
Time bins of 1 s were adopted for all bursts. 
Light curves were rescaled in time to the trigger time of the GRB. 

Fig. \ref{lc} shows the burst light curves. 
In the case of GRB 081007 the 
data present in the GBM catalogue 081007(224) refer to 
GRB 081007B which had very low statistics (Bissaldi et al. 2008). 
We could not find the data directory of GRB 081007 which 
triggered \sw\ (Baumgartner et al. 2008) and \fermi\ (Bissaldi et al. 2008) 121 s after GRB 081007B. 
We only report in Tab. \ref{tab1} the time integrated spectral results 
of GRB 081007 (Bissaldi et al. 2008) . 
We note also that \fermi\ and \sw\ both detected a GRB on November 
18 2008. These are, however, two events that do not 
coincide both spatially and temporally: RA=82.6$^{\circ}$ 
and dec=-43.3$^{\circ}$ is the location (with an uncertainty of 1.6' - Palmer et al. 2008) of the event detected 
by \sw\ at 14:56:36 UT  (Hoversten et al., 2008) while the \fermi\ event is located 
at RA=54$^{\circ}$ and dec=-50.4$^{\circ}$ (with an uncertainty of 2.9$^{\circ}$) and was detected at  21:00:53.5 UT  (Bhat et al., 2008a). 
Only for the \sw\ GRB the redshift is measured ($z$=2.58 D'Elia et al. 2008, no \epo\ measured).

%_____________________________________________________________
%                                             Two column Table 
%_____________________________________________________________
%
\begin{table*}
\caption{Correlation analysis results. $E_{\rm peak}$ and $E_{\rm iso}$ or $L_{\rm iso}$ are normalized to 100 keV and 10$^{52}$ erg respectively. }
\label{tab2}      
\centering          
\begin{tabular}{l c c l}     % 7 columns 
\hline\hline       
  	&	$K$	& 	$\delta$	& $P$ \\ 
$\log{E_{\rm peak,2}}=K+\delta \log{E_{\rm iso,52}}$  &         &          &        \\	  
\hspace{0.2truecm} {\it 100 Pre--Fermi}                      &  0.136$\pm$0.028       &   0.475$\pm$0.027       &  1.4e-24      \\
\hspace{0.2truecm} {\it 10 Fermi}                                  &   0.162$\pm$0.085      &   0.476$\pm$0.079       &  0.004      \\
\hspace{0.2truecm} {\it 110 total}                                 &     0.138$\pm$0.026    &    0.476$\pm$0.025      &    1.8e-27    \\
\hline
$\log{E_{\rm peak,2}}=K+\delta \log{L_{\rm iso,52}}$  &         &          &        \\	  
\hspace{0.2truecm} {\it 100 Pre--Fermi} 	                    &     0.443$\pm$0.029    &   0.391$\pm$0.026       &  2.3e-22      \\
\hspace{0.2truecm} {\it 10 Fermi}  		                    &    0.507$\pm$0.092     &   0.416$\pm$0.080       &  0.006      \\
\hspace{0.2truecm} {\it 110 total}                                 &     0.448$\pm$0.028    &    0.395$\pm$0.024      &    2.4e-25    \\
\hspace{0.2truecm} {\it 51 time resolved spectra}        &   0.599$\pm$0.051      &   0.366$\pm$0.055       &    1.6e-6    \\
\hspace{0.2truecm} {\it 090424 1st peak}                     &  0.348$\pm$0.057       &  0.595$\pm$0.084        &   5.21e-9     \\
\hspace{0.2truecm} {\it 090424 2nd peak}                    &  0.384$\pm$0.043       &  0.578$\pm$0.065        &   0.007     \\
\hspace{0.2truecm} {\it 090618 1st peak}                     &  0.579$\pm$0.052       &  0.442$\pm$0.087        &   0.002     \\
\hspace{0.2truecm} {\it 090618 2nd peak}                    &  0.491$\pm$0.029       &  0.618$\pm$0.043        &  3.3e-5      \\
\hline                    
\hline                  
\end{tabular}
\end{table*}
%----------------------------------------------------------------------------------------------------

% ---------------------------------------------------------------------------------------------------------- 
\begin{figure*}
\centering
\includegraphics[width=17cm]{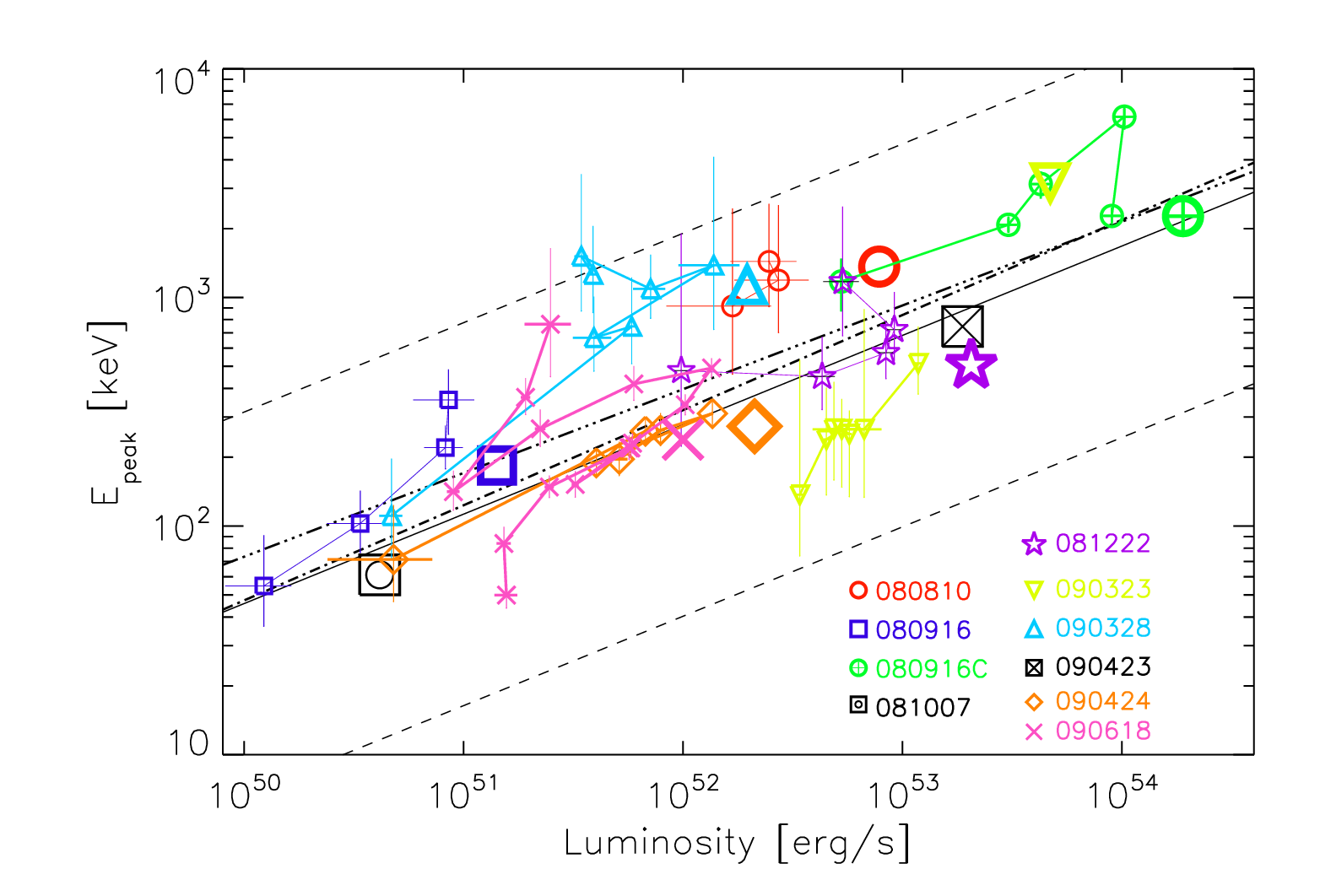}
\caption{
Time resolved \yone\ correlation of \fermi\ GRBs with known redshift. 
Small size symbols show the evolution of \ep\ vs \liso. 
Large size symbols are the location in the 
\yone\ plane of the corresponding bursts when the time 
integrated spectra (Tab. \ref{tab1}) are considered. 
For GRB 081007 and GRB 090423 only the 
time integrated spectrum is available. 
The solid and dotted lines represent the \yone\ correlation 
and its 3$\sigma$ scatter, respectively, as obtained with the 
pre--\fermi\ GRBs (Ghirlanda et al. 2009). The dot--dashed line represents the fit to the 10 \fermi\ GRBs 
(time integrated) and the triple--dot--dashed line is the fit to the 51 time resolved spectra.
}
\label{fg2}
\end{figure*}
%-----------------------------------------------------------------

% ---------------------------------------------------------------
\begin{figure*}
   \centering
   \includegraphics[width=9cm]{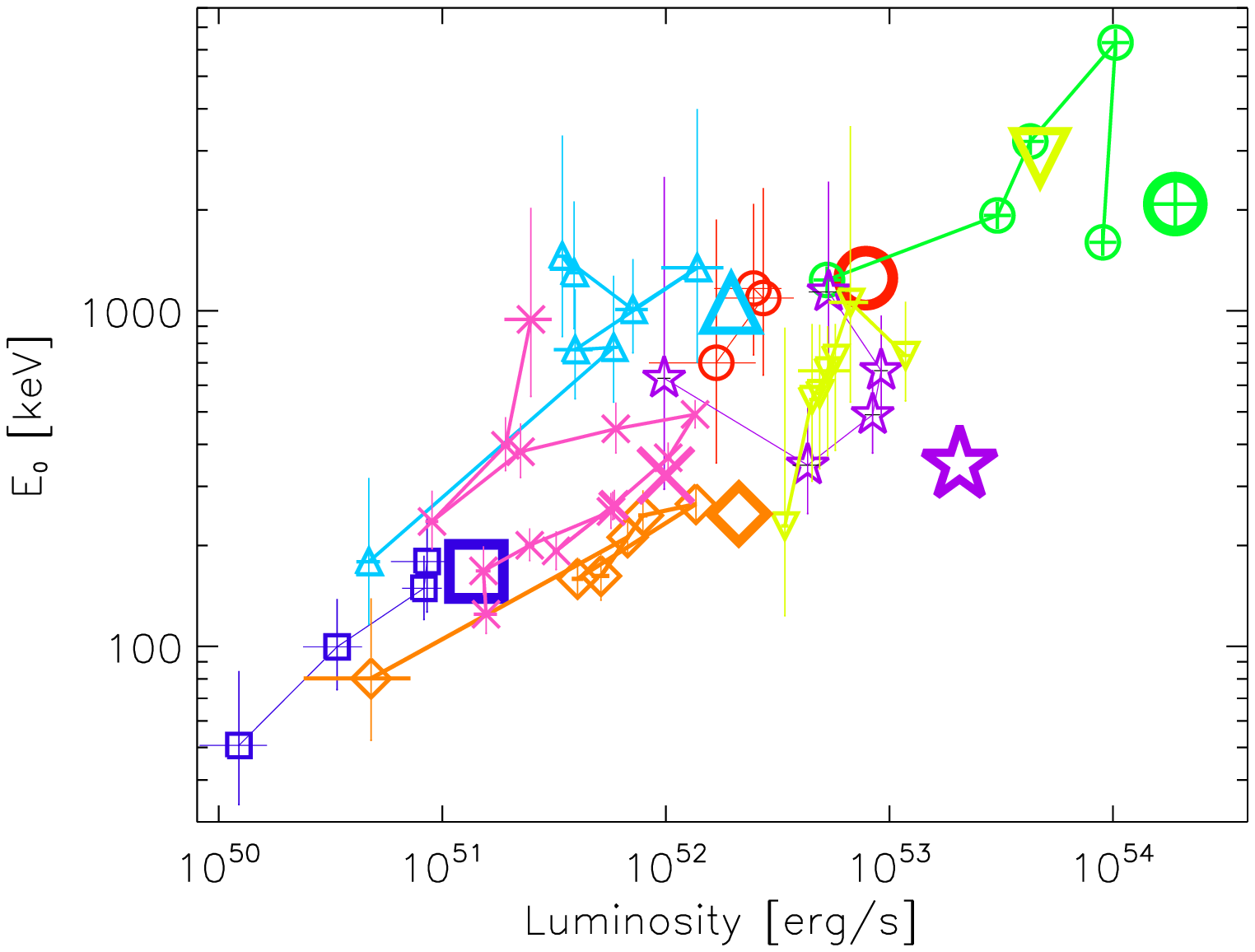}
   \includegraphics[width=9cm]{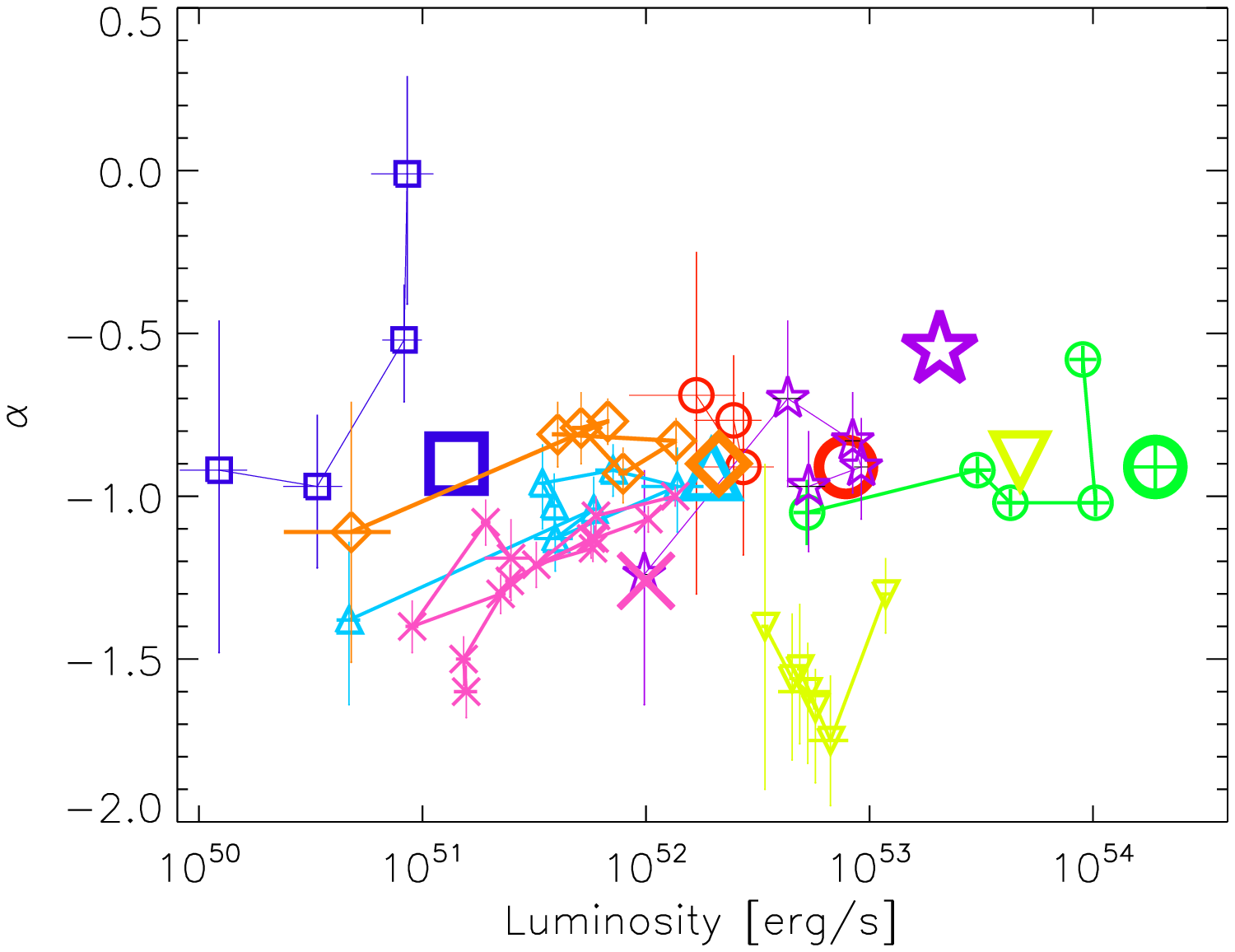}
\caption{
Left panel:
the rest frame energy $E_0=E_{\rm peak}/(2+\alpha)$ as a function of
luminosity for the time resolved spectra and for the 
corresponding time integrated spectra (same symbols as in Fig. \ref{fg2}).  
Right panel: the low energy spectral index $\alpha$ versus the luminosity \liso. 
}
\label{fg3}
\end{figure*}
%------------------------------------------------------------------

For the scope of this work we do not subtract the background 
count rate to the light curves, they are only used here to 
show the time intervals within single GRBs selected for the 
extraction of time--resolved spectra. For 3 GRBs (080905, 080928 and 090423) 
the count rate is too small to perform a time resolved spectral analysis. 
In all the other cases (except GRB 081007) we could divide 
the light curve into time intervals as indicated by the dotted vertical 
lines of Fig. \ref{lc}. 
For GRB 080916C  
we used the time resolved spectral analysis results reported by Abdo 
et al. (2009) who combined the GBM and the LAT data. 
Spectra of the triggered detectors were extracted over the selected time 
intervals defined with the {\it gtbindef} tool. 
The spectrum of the background was extracted in a time interval after 
the burst in order to limit the GRB contamination. 
Rebinning (with the {\it grppha(v3.0.1)} tool) 
was applied to each spectrum in order to have a minimum of 
40 counts per energy channel. 

The response files corresponding to each detector were used for the spectral fitting. 
Spectra were analysed with {\it Xspec(v12)} in the range 8 keV to $\sim$1 MeV.
For most GRBs the spectra of the two (or more) NaI triggered 
detectors were jointly fitted with a cutoff--power law model (CPL)
of the form $F(E)=E^{\alpha}\exp(-E/E_0)$ with a free normalization constant for 
the spectra of the two or more detectors jointly fitted. Most of the fits give for this constant a value close to 1.
% This model consists of a low energy power law with photon 
% index $\alpha$ ending with an exponential cutoff at high 
% energies with characteristic (e--folding energy) $E_{0}$. 
The CPL has been widely used to fit the spectra of GRBs 
and in particular the time resolved spectra (Preece et al. 2000; 
Ghirlanda, Celotti \& Ghisellini, 2002; Kaneko et al. 2006). 
We only analysed the spectra of the NaI detectors because in
most bursts the inspection of the BGO light curve did not 
show any evident signal. Morever, 
%Given the still preliminary calibration of the BGO detectors (REF), 
%we limit the present analysis to the NaI detectors. 
the lower sensitivity of the  BGO detectors with respect to the NaI ones 
would lead, in extracting simultaneous spectra for a joint fit, to a much smaller number of time resolved 
spectra.

For 8 GRBs of Tab. \ref{tab1} we performed a time resolved spectral analysis. 
Note that also the \ba\ time resolved spectra are often fitted 
with the CPL function (e.g. Kaneko et al. 2006; Nava et al. 2008). 
For the purpose of comparing the spectral evolution of 
different bursts in the \yone\ plane the use of the same 
spectral model ensures that the possible biases, e.g. 
the overestimate of the peak energy with respect to the Band model 
(Band et al. 1993), is a common systematic effect of all the 
analyzed spectra (e.g. see Kaneko et al. 2006).  In any case, we also verified if our time resolved 
spectra could be consistent with the Band model finding that, in most cases, we could not constrain the 
high energy power law slope of this model. This also motivated us to to choose the minimal simplest model, i.e. 
the CPL.

We computed the isotropic luminosity of each time resolved spectrum 
by integrating the best fit spectral shape over the rest frame 1 keV--10 MeV energy range. 
Table \ref{tab3} reports the results of the  time resolved analysis: Col. 2 and Col. 3 give
the start and stop times of the time resolved spectra, Col. 4 and Col. 5 give  
the photon spectral index $\alpha$ and the characteristic 
energy $E_0$ (with their 90\% significance errors) respectively.

\section{Results}

\subsection{Evolutionary tracks}

Fig. \ref{fg2} shows the evolutionary tracks of the 8 \fermi\ GRBs with redshift for which time resolved spectral analysis was possible. The number of time resolved spectra extracted per burst depends on its total fluence. 
Our guideline in defining the time intervals was a trade--off between the need to follow the 
rise and decay phases of the single pulses within the light 
curve and to have enough signal in each time resolved spectrum 
 to constrain its spectral parameters. 
In two bursts (GRB 090424 and GRB 090618) a more dense sampling of the 
light curve is possible given their large fluence. 
Here we have limited the number of time bins to have a 
more neat evolutionary track in the \yone\ plane, while in the next sub--section
we will discuss their time evolution in full detail.
For GRB 090323 the time resolved 
spectra up to $\sim$ 100 s after the trigger are best fitted by a 
simple power law model, while only after 100 s a curved model (CPL) 
can be fitted and the value of the peak energy can be constrained. 
Therefore, for GRB 090323 we show in Fig. \ref{fg2} only the spectral 
evolution of the final peaks (those after 100 s in the corresponding 
light curve of Fig. \ref{lc}). 

Fig. \ref{fg2} shows that the prompt  
spectrum evolves in a well defined way, and that \ep\ and $L_{\rm iso}$
are correlated.
The {\it entire} evolutionary tracks  of {\it all} the 8 bursts we
studied lie within the 3$\sigma$ stripe  of the scatter of the \yone\ correlation
defined by the time integrated spectra of different bursts.
Fig. \ref{fg2} shows data points that are associated both to
the rising or descending part of pulses, although they are difficult
to distinguish. In any case, there appears to be no difference between the rising 
and decaying part of the pulses, and this important point will be discussed 
in more detail below.

Since \ep\ is a derived quantity, being $=E_{\rm peak}=E_0 (2+\alpha)$,
it is interesting to verify if the \yonet\ correlation is the result
of underlying correlations of $E_0$ and/or $\alpha$ with $L_{\rm iso}$.
The left panel of Fig. \ref{fg3} shows the rest frame energy 
$E_0=E_{\rm peak}/(2+\alpha)$ as a function of the 
luminosity for the time resolved spectra and for the 
corresponding time integrated spectra (same symbols as in Fig. \ref{fg2}).  
The right panel of Fig. \ref{fg3} shows the low energy spectral 
index $\alpha$ versus the luminosity \liso. 
We can see that $\alpha$ shows no correlation with $L_{\rm iso}$
when considering the ensemble of bursts (but see below the case of
the single pulse of GRB 090618), while $E_0$ does.
We can conclude that it is indeed $E_{\rm peak}$ (or $E_0$, i.e. the e-folding energy of the 
CPL model) the spectral parameter correlating with $L_{\rm iso}$. From the 
right panel of Fig. \ref{fg3} we also see that the low energy photon 
index of the time integrated and time resolved spectra of the analyzed \fermi\ GRBs are consistent with the 
-2/3 limit predicted by synchrotron emission  while they violate the -3/2 synchrotron limit for fast cooling electrons 
(Ghisellini, Celotti \& Lazzati 2000).

We analyzed the \yonet\ correlations obtained with the time resolved spectra. In
Table \ref{tab2} we report the normalization and slope ($K$ and $\delta$, respectively) 
and the chance probability $P$ of the correlation. In total we have 51 time resolved spectra for the 8 GRBs. They define a correlation in the \yone\ plane with a slope 0.37$\pm$0.05 
(triple--dot--dashed line in Fig. \ref{fg2}) consistent with the slope of the correlation defined 
with the time integrated spectra (i.e. 0.39$\pm$0.03 for the 100 pre--\fermi\ GRBs -- solid line in Fig.\ref{fg2} -- or 0.40$\pm$0.02 
for the total of 110 GRBs including the \fermi\ events -- dot--dashed line in Fig. \ref{fg2}).

% ----------------------------------------------------------------------------------
\begin{figure*}
   \centering
   \includegraphics[width=9cm]{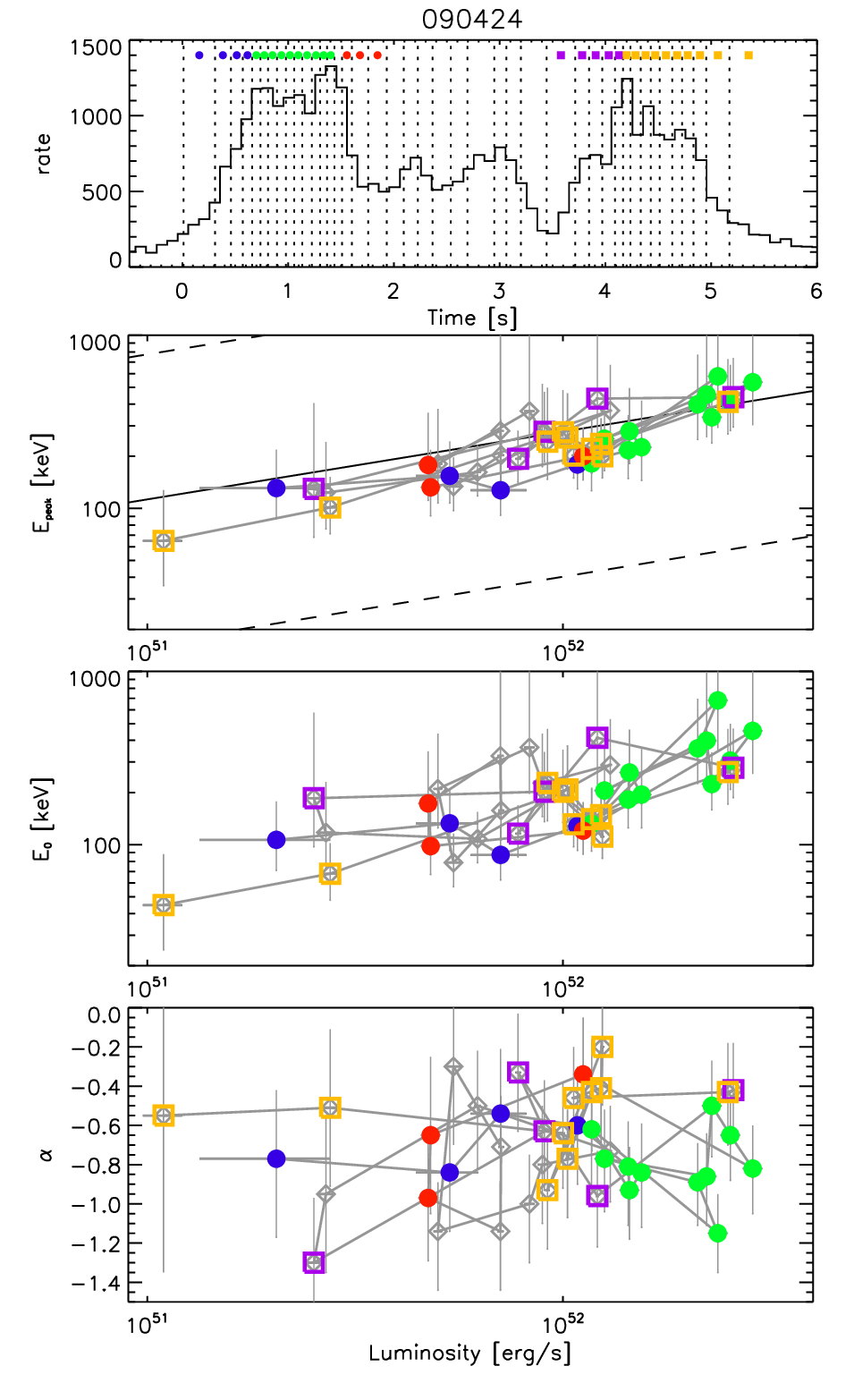}
   \includegraphics[width=9cm]{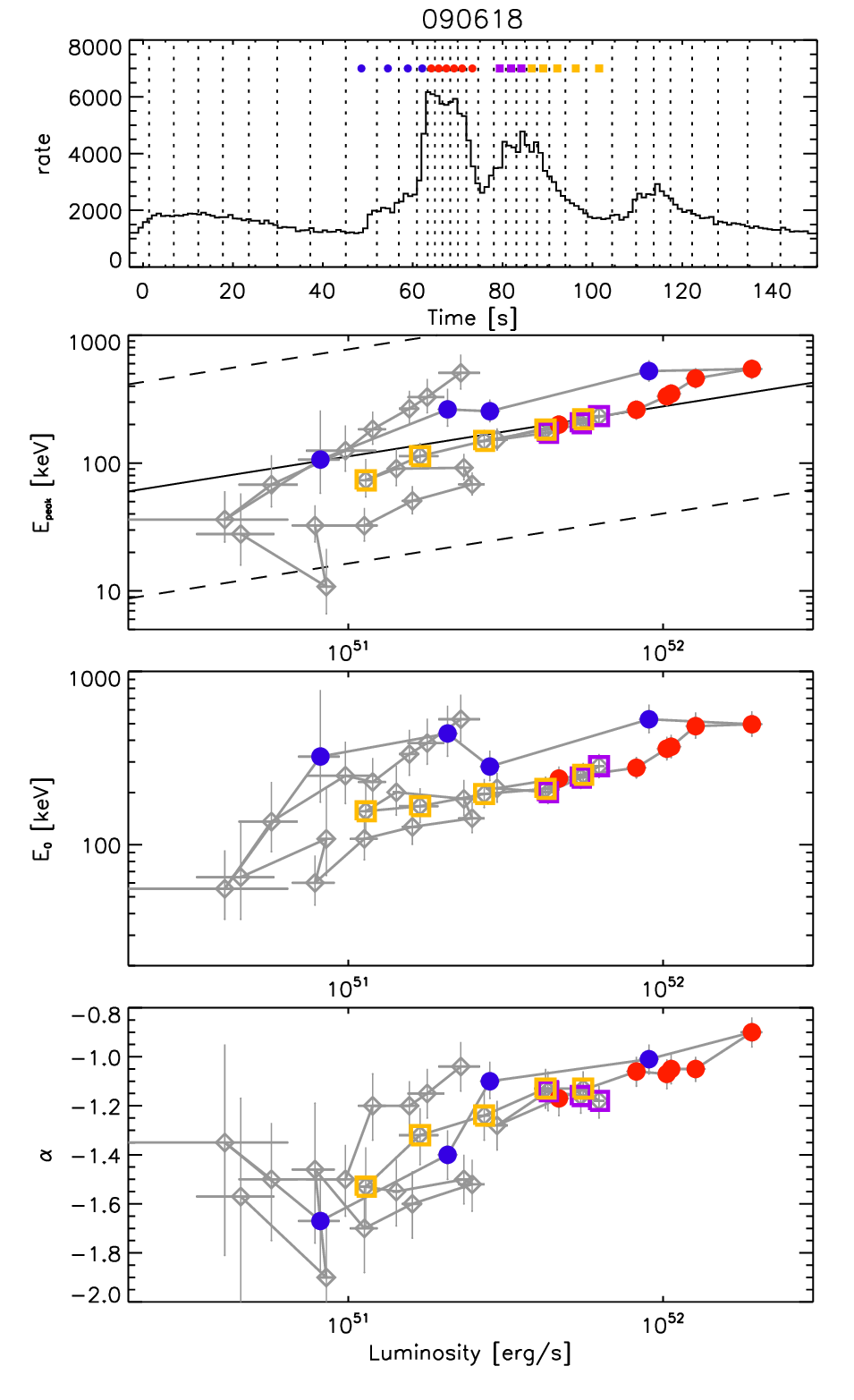}
\caption{
High time resolution spectral analysis of 
GRB 090424 (left) and of GRB 090618 (right). 
Top: light curve (0.1s resolution) and time intervals 
of the extracted spectra (vertical dotted lines). 
The coloured symbols represent the spectra highlighted in the 
panels below. 
Spectral evolution panels (from top to bottom): 
rest frame peak energy \ep, rest frame characteristic energy 
$E_0$ and photon index $\alpha$ versus isotropic luminosity. 
The filled circles and the open squares identify the first 
and the second peak  of the light curves. 
The colour code marks the rise/decay phase within each peak.  
}
\label{fg4}
\end{figure*}
%-----------------------------------------------------------------------------------

\subsection{Rising and decaying phases of single pulses} 

GRB 090424 consists of two peaks overlapped with a total 
duration of 6 seconds. 
GRB 090618 has a complex light curve made of a smooth precursor 
followed by two intense peaks partially overlapped (see Fig. \ref{lc}) 
for a total duration of 150 s. 
Their high count rates allow a time resolved spectral analysis
with a dense time sampling.
These two GRBs are well suited to study how the spectrum evolves 
during the rise and decay phases of their pulses.

We divided the time interval of the duration of GRB 090424 so 
that each extracted spectrum had a signal--to--noise ratio S/N$\geq$30, 
integrated over the 8 keV--1 MeV energy range. 
This gives a total of 42 spectra distributed in the  
$\sim$6 s of duration of GRB 090424. 
For GRB 090618 we required a S/N ratio of 50 and obtained 34 
time resolved spectra in the 150 s of its duration. 
These spectra were extracted and analysed as described in \S 3. 
In Fig. \ref{fg4} the light curve (with 0.1s time resolution) and the three panels of 
the correlation between the spectral parameters 
(\ep, $E_0$ and $\alpha$) and the luminosity \liso\ are shown for GRB 090424 (left) 
and GRB 090618 (right). 
In these plots we have marked with different symbols the spectral 
evolution of the different peaks identified in the light curve and 
with different colours the rise and decay phase of single pulses. 
The different peaks have also been fitted individually in the \yone\ plane (second panels of Fig. \ref{fg4}) 
and the results are shown in Table \ref{tab2}. We note that these 4 peaks (two of GRB~090424 and two of GRB~090618) 
define a very tight \yone\ correlation with slope between 0.45 and 0.6. 

These plots indicate that the {\it rising and decaying phases are 
indistinguishable in a \yone\ plot}. 
Intriguingly, we also find that for GRB 090618 there is a correlation
between $\alpha$ and $L_{\rm iso}$.
Since also $E_0$ correlates with $L_{\rm iso}$ (as for the other bursts),
this results in a even tighter \yonet\  correlation for the two pulses of this GRB.

\section{Discussion \& Conclusions}

The largest sample of long GRBs with measured redshift and \epo\ collected recently (e.g. Ghirlanda et al. 2009; Nava et al. 2008) defines a strong correlation $E_{\rm peak}\propto L_{\rm iso}^{0.4}$ with scatter $\sigma \simeq 0.26$. A similar strong correlation exists between the peak energy and the isotropic energy, i.e. \ep$\propto E_{\rm iso}^{0.5}$.  The time integrated spectra of the 8 \fermi\ GRBs with measured redshift (open symbols in Fig. \ref{corr_upd}) are consistent with both the \ama\ and the \yone\ correlation defined by 100 pre--\fermi\ bursts (grey filled circles in Fig. \ref{corr_upd}). In the \yone\ correlation, \ep\ is that of the time integrated spectrum and \liso\ is computed (as done by Yonetoku et al. 2004) using the value of \ep\ of the time integrated spectrum and the peak flux of the GRB. This luminosity is somehow different from the luminosity of the brightest phase of the burst, i.e. the luminosity of the peak spectrum. A discussion about these two definitions of \liso\ can be found in Ghirlanda et al. (2005). 

\subsection{Selection effects}
The sample of bursts defining the \ama\ and \yone\ correlations is heterogeneous: Nava et al. (2008) considered 83 pre--\fermi\ GRBs detected by different instruments (\ba, \sax, \he, \ko\ and \sw) since 1997, a more recent update (Ghirlanda et al. 2009) consider 100 GRBs.  Here we have added the 8 \fermi--GRBs with measured redshift. Adding new GRBs to the above correlations does not represent a secure test of their physical nature especially when there is the suspect that these correlations are the result of instrumental selection effects. Different instrumental selection effects could be biasing the samples of bursts used to define the \ama\ and \yone\ correlations. Butler et al. (2007) (see also Shahmoradi \& Nemiroff 2009) argued that the spectral--energy  correlations, in particular the \ama\ and the \yone\ defined by the time integrated spectra, are the effect of the trigger threshold, therefore having no physical relevance for the understanding of the GRB emission mechanism. In Ghirlanda et al. (2008) we investigated this issue by studying instrumental selection effects possibly biasing a sample of 76 bursts (updated to Sep. 2007) with measured redshifts. This sample defines strong correlations in the observer frame \epof\ and \epop\ planes ($F$ and $P$ are the fluence and peak flux) where the instrumental selection effects can be studied. Two selection effects were considered for the detectors on-board \ba, \sax\ and \sw: the trigger threshold, i.e. the minimum flux a burst must have to trigger a given detector, and the spectral threshold, i.e. the minimum fluence a burst must have in order to constrain its spectral parameters (in particular the peak energy \epo). Our results indicate that: (i) both the selection effects are functions of \epo\ but the spectral threshold is dominating over the trigger threshold, implying that the \ama\ or \yone\ correlations are not due to the trigger threshold; (ii) the \sw\ spectral threshold is biasing the \sw\ GRB sample with redshift added to the \ama\ correlation in the last three years; (iii) selection effects are present but they do not determine the spectral--energy correlations. 

Another way to test the incidence of selection effects on the \ama\ and \yone\ correlation is to verify how much the slope, normalization and scatter change by separating the heterogenous sample of GRBs with redshift into sub-samples of bursts detected by different instrument (e.g. Butler et al. 2007). Butler et al. 2007 found similar slopes but different normalizations of the \ama\ correlation by considering the pre--\sw\ and the \sw\ sample (but see Amati et al. 2009). Unfortunately, \sw\ bursts have a very narrow range of \epo\ limiting the robustness of this test (Nava et al. 2008). Furthermore, different instruments like  \ba\ and \sax\ can have very similar detector thresholds and bias, in a similar way, the GRB samples that they detect. 

\subsection{Time resolved \yonet\ correlation}

All  the spectral--energy correlations have been derived considering the time integrated GRB spectral properties. 
%Liang et al. (2004), however, showed that, within individual \ba\ GRBs, there is a strong correlation between the peak energy and the flux of time resolved spectra. Unfortunately, the lack of any measured redshift for those  \ba\ GRBs  prevented a direct comparison of their evolutionary tracks in the \yone\ plane. Firmani et al. (2009) presented the first evidence of the existence of a time resolved \yonet\ correlation within \sw\ GRBs with measured redshifts. This correlation is similar to that defined by time integrated spectra. 
%Similar results were found recently by Krimm et al. (2009) who verified that  the \ama\ correlation holds among different pulses within a burst. In their work they analyzed the bursts detected simultaneously by \sw\ and {\it Suzaku}. The larger energy range of the combined spectra allows to derive better constraints on the peak energy of time resolved spectra. 
By studying the spectral evolution of the 8 \fermi\ GRBs with measured redshift we also find that a correlation \yonet, between the rest frame peak energy \ep\ and the bolometric isotropic luminosity, exists within individual bursts (Fig. \ref{fg2}). This \yonet\ correlation can also extend over two orders of magnitude in both \ep\ and \liso\ within the duration of a burst. 
The evolutionary tracks defined by the 8 \fermi\ GRBs lie in the upper part of the \yone\ correlation. This could be due to a systematic underestimate of the luminosity in time resolved spectra with respect to time integrated spectra (which are used to define the \yone\ correlation plotted as a solid line in Fig. \ref{fg2}). Indeed, the time resolved spectra are more frequently fitted with a CPL model which lacks the high energy power law component of time integrated spectra. 

The finding of a \yonet\ correlation within individual GRBs,  consistent with the \yone\ correlation defined by time integrated spectra, is the strongest argument in favour of a physical origin of this correlation and the strongest argument against instrumental selection effects biasing the observed correlations. 

\subsection{Interpretations of the spectral--energy correlations}

A convincing way to ensure the reality of the spectral--energy correlations would be to find a robust physical interpretation. 
The proposed interpretations of the \yone\ and \ama\ correlations can be divided into two classes: (a) kinematic interpretations in which the link between \ep\ and \liso\ is established by the configuration of the emission region, i.e. a uniform jet observed at different angles (Yamazaki et al. 2004), an inhomogeneous jet model (e.g. Nakamura 2000; Kumar \& Piran 2000) made up of multiple sub--jets or emission patches (Toma et al. 2005) or a ring--shaped emission region (Eichler \& Levinson 2004); (b) radiative interpretations in which it is the emission mechanism of the prompt phase to link \ep\ and \liso\ as in the case of a spectrum dominated by a thermal component (Meszaros \& Rees 2007; Ryde et al. 2006; Thompson, Meszaros \& Rees 2007), in the case of  photospheric emission dominated by magnetic reconnection (Giannios \& Spruit, 2007) or when the emission is synchrotron radiation from the external shock  (Panaitescu et al. 2009).  

The common feature of the kinematic models in reproducing the \yone\ or \ama\ correlation is the viewing angle under which different GRBs are observed. Both the off-axis and the sub--jet models need  to assume the existence of a on--axis  correlation between the peak energy and the luminosity, whose origin could be instead related to the radiative process.  Indeed, the kinematical models that (under some assumptions about the typical jet opening angle distribution) succeed in reproducing \ep $\propto$ $L_{\rm iso}^{0.5}$ should still explain a similar correlation within individual GRBs, i.e. a time dependent correlation \yonet, which can extend over 2 orders of magnitude (e.g. Fig. \ref{fg3}).

The simplest way to explain the \yone\ correlation is to assume that only the bulk Lorentz factor $\Gamma$ changes.
Since \ep$\propto\Gamma$ and $L\propto \Gamma^2$, we recover \ep$\propto  L^{1/2}$. But this assumes that, in the comoving frame, both \ep$^\prime$ and $L^\prime$ are the same {\it even if different} $\Gamma$--factors are required, and this seems unlikely (both when considering the \yone\ correlation defined by different bursts or the \yonet\ correlation holding within individual GRBs).

If the emission is due to the synchrotron process, the peak frequency \ep$\propto B\Gamma\gamma_{\rm p}^2$ (where $\gamma_{\rm p}$ is the random Lorentz factor of the electrons emitting at the peak) and  $L\propto N B^2\Gamma^2 \gamma_{\rm p}^2$, where $N$ is the  number of the electrons having $\gamma_{\rm p}$. Therefore a change of the quantity $\Gamma B$, maintaining the same $\gamma_{\rm p}$ and $N$, would give \ep$\propto  L^{1/2}$. But the prompt emission almost surely occurs in the fast cooling regime,  implying that the resulting synchrotron spectrum cannot be harder than $L(E)\propto E^{-1/2}$ (Ghisellini et al. 2000) while we observe (e.g. Preece et al. 2000; Ghirlanda, Celotti \& Ghisellini 2003) also in the \fermi\ GRBs  
(Fig. \ref{fg3} right panel) much harder spectra. Furthermore, it is seems hard to maintain the same $N$ and $\gamma_{\rm p}$ while changing $\Gamma B$.

Quasi--thermal Comptonization could well explain the fact, found in GRB 090618, that both $E_0$ and $\alpha$ correlate with $L$.
In fact, if the seed photons for Compton scattering remain the same, an increase of the plasma temperature would increase the Comptonization parameter  $y\sim 4\tau KT_{\rm e}/ (m_{\rm e} c^2)$, producing both an harder spectrum  and a larger \ep\ ($\tau$ is the optical depth of scattering electrons, and $T_{\rm e}$ their temperature). On the other hand, for likely bulk Lorentz factors $\Gamma\sim$10$^2$--10$^3$, the comoving temperature is below 1 keV, implying $\tau>10^3$ to reach the required $y\sim 10$, needed to account for the observed flat spectra. With this values of $\tau$ the resulting spectrum would saturate to a Wien--like
spectrum, not to a cutoff power law. A very large value of $\tau$ would also lengthen any variability timescale.

It has been suggested (e.g. Borgonovo \& Ryde 2001; Ryde \& Petrosian 2002) that the off--latitude emission that follows an abrupt switch--off of the fireball introduces a spectral--energy dependence, since the observer sees progressively less beamed (and less blue--shifted) emission. However, this could explain {\it only the decaying phase} of the pulse. In this paper we analyzed the spectral evolution of two of the most intense bursts in our sample: they allow to make a dense sampling of their light curves in order to extract time resolved spectra. These are GRB 090424 and GRB 090618 (Fig. \ref{fg3} right and left panel, respectively).   Our findings indicate clearly that there is the same \yonet\ correlation during the rise and the decay phase of different pulses within these two GRBs. The time evolution is so that during the rise phase both \ep\ and \liso\ increase to the maximum value and during the decay they decrease along the same evolutionary track that they followed during the rise phase. In the case of GRB 090618 these considerations are valid also for the correlation between $\alpha$ and \liso (bottom right panel of Fig. \ref{fg3}). 

In the attempt to explain the \ama\ correlation, Thompson Meszaros \& Rees (2007) pointed out the importance of shear layers shocks to extract a large fraction of the bulk kinetic energy of the fireball, leading to a black body spectrum. The same arguments could be used to explain the \yone\ correlation, in different bursts and the \yonet\ correlation within individual bursts as well. The problem with this interpretation is that one of the key assumption of their scenario is that the value of the bulk Lorentz factor in the dissipation region must be fine tuned (it must be of the order of $1/\theta_{\rm j}$, where $\theta_{\rm j}$ is the opening angle of the fireball). This assumption is relaxed in the ``reborn fireball" scenario  (Ghisellini et al. 2007), but there remains to be explained why so few bursts have pure black body spectra (Ghirlanda, et al. 2003), and, even when adding a power law component (Ryde et al. 2005), its slope is too soft to explain low energy data (in the keV band), as shown by Ghirlanda et al. (2007).

We can conclude that new ideas are called for explaining what emerges to be a general and well defined property of the prompt emission of GRBs.

%Our results also indicate that, although \epo=$E_{0}(\alpha+2)$ (where $E_{0}$ and $\alpha$ are the free parameters of the CPL model adopted to fit the time resolved spectra), we found a strong correlation between the luminosity and the (rest frame) $E_0$ in almost all bursts (Fig. \ref{fg3}). Instead we do not find a clear correlation between the luminosity and the low energy spectral index $\alpha$. Finally our analysis confirms that also the \fermi\ time resolved spectra are harder (Fig.\ref{fg3}) than the -3/2 synchrotron limit of fast cooling electrons (Ghisellini et al. 2000), as already shown for several other GRBs detected by different instruments (e.g. Preece et al. 2000; Kaneko et al. 2006, Sakamoto et al. 2006).

%Motivated by the above results 
%\section{Conclusions}

%Our results clearly indicate that there must be a physical mechanism operating during the prompt GRB emission which links the peak energy \ep and the luminosity \liso.   The \yone\ correlation defined by  different GRBs is not due to selection effects, 
%but instead has a physical origin.  The mechanism responsible for the \yonet\ correlation (i) must be very robust, to work during the prompt emission phase of individual bursts as well as in different bursts; (ii) it is very likely to be due to the radiative process of the GRB prompt emission phase. 

\begin{acknowledgements}
We are grateful to Z. Bosnjak, D. Burlon, A. Celotti, C. Firmani, M. Nardini and F. Tavecchio 
for stimulating discussion. ASI is thanked for grant I/088/06/0.  A PRIN-INAF grant is acknowledged for 
funding. This research made use of the \fermi--Gamma Burst Monitor data 
publicly available via the NASA-HEASARC data center.  
\end{acknowledgements}

\section{Appendix}

%_____________________________________________________________
%
\begin{table*}
\caption{Spectral results of the time resolved analysis.}
\label{tab3}      
\centering          
\begin{tabular}{l l l l l l l l l l}     % 7 columns 
\hline\hline       
   GRB & t$_1$ & t$_2$ & $\alpha$ & $E_{\rm 0}$  & $\beta$  &  $\chi^2$(dof) &  $F_{-6}$ \\
   	   &   s      &   s       &                &    keV             &           &                       & erg/cm$^2$ s \\
\hline                    
     	   	     &         &      &                                         &                                   &            &                 &           \\
080810(549) & -10 & 20	& -0.77$_{-0.24}^{+0.2}$   & 268$_{-98}^{+211}$   &    & 214(215)  & 0.24 \\
		     &  20 & 30	& -0.91$_{-0.27}^{+0.23}$ & 251$_{-103}^{+283}$ &   & 185(168)  & 0.27 \\
		     &   40 & 46	& -0.69$_{-0.61}^{+0.44}$ & 161$_{-80}^{+269}$   &   & 139(140)  & 0.16 \\
		     &  46	& 56 & -1.76$_{-0.18}^{+0.2}$	  &                             &  & 184(164)  & 0.21	\\
     	   	     &         &      &                                         &                                   &                                  &                 &  \\
080916(009) &0.004&3.58&-0.58$\pm$0.04               & 310$\pm$19              &   -2.63$\pm$0.12  &            &       \\
     	   	     & 3.58 &7.68& -1.02$\pm$0.02              &  1193$\pm$142          &   -2.21$\pm$0.03&              & \\
		     & 7.68&15.87&-1.02$\pm$0.04              & 602$\pm$82               &  -2.16 $\pm$0.03  &              &\\
		     &15.87&54.78&-0.92$\pm$0.03             &370$\pm$24                &   -2.22 $\pm$0.02  &             & \\
		     &54.78&100.86&-1.05$\pm$0.10           &242$\pm$60               &    -2.16$\pm$0.05 &               &\\
		     &          &          &                                     &                                    &                                  &                      &     \\ 
080916(406) & -2 	& 2	& -0.01$_{-0.4}^{+0.3}$	  & 106$_{-31}^{+38}$      &   & 	122(131)	& 0.42  \\
		     & 2	& 8	& -0.52$_{-0.19}^{+0.17}$ & 88.2$_{-17}^{22}$        &     &  156(148)	& 0.4    \\
		     & 12	& 24 & -0.97$_{-0.25}^{+ 0.22}$ & 59$_{-15}^{+23}$	&   &  164(175)	& 0.17 \\
		     & 24	& 50	& -0.92$_{-0.56}^{+ 0.46}$ & 30$_{-10}^{+20}$       &     & 	225(209) 	& 0.06 \\
     	   	     &         &      &                                         &                                   &                                  &                 &  \\
081222(204) & -2    &  2	& -0.97$_{-0.2}^{+0.17}$  & 302$_{-127}^{+341}$  &  			 &	73(70)   & 	0.54 \\
		     & 2	& 4	& -0.91$_{-0.16}^{+0.15}$ & 176$_{-47}^{+81}$ 	& 				 &     55(59)    & 	1.39	\\
		     & 4 	& 6	& -0.83$_{-0.16}^{+0.15}$ & 130$_{-30}^{+47}$     &			         &	68(60)    & 	1.29	\\
		     & 6 	& 8	& -0.7$_{-0.27}^{+0.24}$	  & 92$_{-26}^{+45}$ 	&			& 	61(50)    &	0.7	\\
		     & 8 	& 20	& -1.24$_{0.4}^{+0.32}$	&   167$_{-89}^{+498}$   & 			 &     82(89)    &	0.15 \\
     	   	     &         &      &                                         &                                   &                                  &                 &           \\
090323(002) & -4.6	& 3.1 & -1.98$_{-0.08}^{+0.08}$  &				&			  &    105(88) & 	0.64	\\
		     & 3.1   & 9.6 & -1.57$_{-0.05}^{+0.06}$  &				& 				& 	103(91)  &	2.84	\\
		     & 9.6   & 15.2&-1.5$_{-0.03}^{+0.03}$   & 				& 			&      108(90)   & 	5.0  \\
		     & 15.2 &  20.9&-1.47$_{-0.03}^{+0.03}$ & 				& 			& 	108(88)	&	5.36	\\
		     & 20.9 & 27.5 & -1.71$_{-0.05}^{+0.05}$ & 				&			&       96(91)	& 	1.83	\\
		     & 27.5 & 34.6 & -1.9$_{-0.06}^{+0.06}$  &					& 			&	123(89)	&	1.0 \\
		    & 34.6	& 41.1 & -1.67$_{-0.05}^{+0.05}$ & 				& 				& 	87(88)	&	2.11	\\
		    & 41.1	& 46.6 & -1.47$_{-0.03}^{+0.03}$  &			&				&	84(90)	&	5.8  \\
		   & 46.6	& 52.3 & -1.5$_{-0.04}^{+0.04}$   &				& 				& 	118(87)	&	4.9	\\
		   & 52.3	& 57.9 & -1.44$_{-0.04}^{+0.03}$ &				& 				&	127(88)	&	6.2  \\
		   & 57.9	& 62.9 & -1.36$_{-0.03}^{+0.03}$ &					& 			&      103(90)    &	9.8	\\
		   & 62.9	& 67.3 & -1.3$_{-0.02}^{+0.03}$   &					& 			&	129(90)	&	14.7 \\
		   & 67.3	& 73    & -1.62$_{-0.03}^{+0.03}$ &					&				&	93(87)	&	3.4	\\
		   & 73	& 80.2 & -1.93$_{-0.07}^{+0.07}$ &				& 				&	105(91)	&	0.83 \\
		   & 80.2	& 87.8 & -2.0$_{-0.07}^{+0.07}$   &					& 			&	144(91)	&	0.66	\\
		   & 87.8	& 95.6 & -2.0$_{-0.08}^{+0.08}$  & 					& 				&	98(90)	&	0.61	\\
		   & 95.6	& 103.2& -1.4$_{-0.5}^{+0.5}$	   & 50$_{-23}^{+145}$& 				&	98(87)	&	0.27	\\
		   & 103.3 & 110.8 & -1.56$_{-0.25}^{+0.2}$&119$_{-51}^{+81}$ &				&	102(87)	&	0.36	\\
		   & 110.8 & 118.3 &	-1.6$_{-0.22}^{+0.15}$& 145$_{-	64}^{+52}$&			&	129(86)	&	0.38	\\
		   &118.3	& 125.5 & -1.65$_{-0.23}^{+0.12}$& 158$_{-	74}^{+42}$& 			&	114(86)    &	0.43	\\
		   &125.5  & 132.9 & -1.53$_{-0.23}^{+0.2}$  & 124$_{-	50}^{+75}$& 			&	99(87)	&	0.38	\\
		   &132.9  & 140    &	-1.75$_{-0.2}^{+0.2}$    & 232$_{-115}^{+546}$& 			&	106(87)	&	0.05 \\
		   &140	& 150    &	-1.3$_{-0.12}^{+0.11}$  & 161$_{-43}^{	+72}$  &			&	135(84)	&	0.9	\\
     	   	     &         &      &                                         &                                   &                                  &                 &           \\
090328(401) & 0	& 8	& -1.03$_{-0.1}^{+0.1}$	& 748$_{-238}^{+472}$  & & 168(181) &	1.6	\\
		     & 8	& 14	&  -0.96$_{-0.14}^{+0.12}$ & 838$_{-355}^{+1080}$ &   & 215(166) &	1.45	\\
		     & 14 	& 18	&  -0.92$_{-0.08}^{+0.08}$ & 581$_{-149}^{+240}$   &  & 167(167) &	3.0  \\
		     & 18	& 20	&  -0.97$_{-0.14}^{+0.12}$ & 774$_{369}^{1000}$    &	&   91(102)    &	1.94	\\
		    &  20 	& 24	&  -1.13$_{-0.1}^{+0.1}$	   & 441$_{-126}^{+227}$  &  & 180(154)   & 1.52 \\
		    &  24 	& 26	&  -1.04$_{-0.1}^{+0.1}$	&  448$_{-140}^{+285}$    &  &  118(119)  & 2.35 \\
		    &	26   &  30	&  -1.6$_{-0.06}^{+0.07}$ &					 & & 144(132)  & 1.54	  \\
		    &  55	& 62	&  -1.38$_{-0.26}^{+0.24}$ & 103$_{-36}^{+80	}$	 &      &  169(159)  & 0.25	  \\
	     &         &      &                                         &                                   &            &                 &           \\
\hline                  
\end{tabular}
\end{table*}

\begin{table*}
\caption{Spectral evolution. Continued }
\label{table:2}      
\centering          
\begin{tabular}{l l l l l l l l l l}     % 7 columns 
\hline\hline       
   GRB & t$_1$ & t$_2$ & $\alpha$ & $E_{\rm 0}$  & $ \beta$   &  $\chi^2$(dof) &  $F_{-6}$ \\
   	   &   s      &   s       &                &    keV             &           &                       & erg/cm$^2$ s \\
\hline    
     	   	     &         &      &                                         &                                   &            &                 &           \\
090424(592) & -1	& 1	& -0.81$_{-0.1}^{+0.1}$	 & 103$_{-13}^{+16}$      &	&	80(70)   &	      3.4	 \\
		     & 1 	& 2	& -0.83$_{-0.07}^{+0.07}$ & 172$_{-20}^{+25}$	& 	&      71(73)   &	      11.4   \\
		     & 2	& 3	& -0.93$_{-0.09}^{+0.08}$ & 159$_{-23}^{+30}$      &	&      70(66)   & 	6.5  \\
		     & 3	& 4	& -0.79$_{-0.11}^{+0.11}$ & 105$_{-16}^{+20}$	& 	&	84(59)   & 	4.4	\\
		     & 4	& 6	& -0.77$_{-0.07}^{+0.07}$ & 137$_{-15}^{+17}$	&	&	104(76) &		5.78 \\
		     & 6	& 10	& -1.11$_{-0.4}^{+0.4}$	&   52$_{-18}^{+38}$	& 	& 	62(62)   &		0.4	\\
     	   	     &         &      &                                         &                                   &            &                 &           \\
090618(353) & 0	& 3	& -1.19$_{-0.13}^{+0.12}$ & 	612$_{-251}^{+704}$	&  &	124(120) & 	2.1	\\
		      & 3	& 14	& -1.08$_{-0.07}^{+0.07}$ &	258$_{-41}^{+55}$	&  &	296(188) &	1.64 \\
		      & 14 	& 40	& -1.4$_{-	0.08}^{+0.08}$  &     153$_{-26}^{+36}$	&  & 267(214) &	0.76	\\
		     &  50	& 60	& -1.3$_{-	0.06}^{+0.06}$  &	247$_{-40}^{+53}$   &      & 246(187) &	1.88	\\
		     & 60	& 63	& -1.06$_{-0.06}^{+0.06}$ & 	289$_{-44}^{+57}$	&  & 166(150) &	5.12	\\
		     & 63	& 67	& -1.0$_{-	0.03}^{+0.03}$  & 	319$_{-28}^{+33}$	& 	& 296(183) &	11.7	\\
		     &	67	& 70	& -1.07$_{-0.04}^{+0.04}$ &	237$_{-23}^{+26}$   &  & 204(167) &	8.9	\\
		     & 70	& 75	& -1.13$_{-0.04}^{+0.04}$ &	172$_{-16}^{+18}$   & 	 & 219(167) &	5.0	\\
		     & 75	& 80	& -1.21$_{-0.07}^{+0.07}$ &	125$_{-15}^{+18}$   &      & 280(256) &	2.76	\\
		     & 80	& 85	& -1.16$_{-0.04}^{+0.04}$ &	170$_{-15}^{+18}$   &  & 203(167) & 	5.0	 \\
		     & 85	& 88	& -1.13$_{-0.06}^{+0.06}$ &	164$_{-18}^{+22}$   &  & 153(144) &	5.0	 \\
		     & 88	& 100& -1.26$_{-0.05}^{+0.05}$ &	130$_{-13}^{+16}$   & & 219(188) & 	2.1	\\
		     & 100 	& 114& -1.5$_{-0.08}^{+0.07}$ & 	109$_{-16}^{+20}$   & & 233(189) & 	1.35	\\
		     & 114 & 130	 & -1.6$_{-0.08}^{+0.08}$ &	81$_{-10}^{+13}$     & 	& 247(193) & 	1.68	 \\
     	   	     &         &      &                                         &                                   &            &                 &           \\

\hline                  
\end{tabular}
\end{table*}

%---------------------------------------------------------------------

\end{document}